\documentstyle[12pt,epsf,rotate]{article}

\def\baselinestretch{1.1}
\catcode`\@=11

\@addtoreset{equation}{section}

\def\@normalsize{\@setsize\normalsize{15pt}\xiipt\@xiipt
           \abovedisplayskip 14pt plus3pt minus3pt%
           \belowdisplayskip \abovedisplayskip
           \abovedisplayshortskip  \z@ plus3pt%
           \belowdisplayshortskip  7pt plus3.5pt minus0pt}
\def\small{\@setsize\small{13.6pt}\xipt\@xipt
           \abovedisplayskip 16pt plus3pt minus3pt%
           \belowdisplayskip \abovedisplayskip
           \abovedisplayshortskip  \z@ plus3pt%
           \belowdisplayshortskip  7pt plus3.5pt minus0pt
             \def\@listi{\parsep 4.5pt plus 2pt minus 1pt
             \itemsep \parsep
             \topsep 9pt plus 3pt minus 3pt}}
\def\underline#1{\relax\ifmmode\@@underline#1\else
        $\@@underline{\hbox{#1}}$\relax\fi}
\@twosidetrue

\long\def\@makecaption#1#2{
 \vskip 10pt \setbox\@tempboxa\hbox{#1: #2}
 \ifdim \wd\@tempboxa >\hsize #1: #2\par \else \hbox
        to\hsize{\box\@tempboxa\hfil} \fi}

\catcode`\@=12

\relax

            \def\dash{\hbox{---}}
\def\tilde{\widetilde}

\def\tr{\mathop{\rm tr}}  \def\Re{\mathop{\rm Re}}
\def\Im{\mathop{\rm Im}}

 \def
\gta {\mathrel{\vcenter {\hbox{$>$}\nointerlineskip\hbox{$\sim$}}}}

\def\gtap{\raisebox{-.4ex}{\rlap{$\sim$}} \raisebox{.4ex}{$>$}}
\def\bar{\overline}

\def\beq{\begin{equation}} \def\eeq{\end{equation}} \def\bea{\begin{eqnarray}}
\def\eea{\end{eqnarray}} \def\bq{\begin{quote}} \def\eq{\end{quote}}  

\relax

\renewcommand{\(}{\left(} \renewcommand{\)}{\right)} \renewcommand{\[}{\left[}
\renewcommand{\]}{\right]} \renewcommand{\bar}[1]{\overline{#1}}

\def\SMo{Standard Model}
\def\SM{\SMo\ } \def\SMp{\SMo.\ } 
\def\EW{electro--weak } 
\def\GBo{Goldstone Boson}
  \def\GBs{\GBo s\ } \def\GBsp{\GBo s.\ }
\def\RG{renormalization group }

\def\sixtpi{\frac{1}{16\pi^{2}}}
\def\lnlambda{\ln \left( \frac{\Lambda^{2}}{\mu^{2}} \right)}
\def\vr2{v_{R}^{2}}
\def\kap2{\kappa^{2}}
\def\kapp2{\kappa'^{2}}
\def\hvr{\hat{v}_{R}}
\def\hvr2{\hat{v}_{R}^{2}}
\def\hkap2{\hat{\kappa}^{2}}
\def\hkapp2{\hat{\kappa'}^{2}}

\begin{document}


\begin{titlepage}
  \renewcommand{\baselinestretch}{1}
  \renewcommand{\thefootnote}{\alph{footnote}}
  \thispagestyle{empty}

\ \vskip -1.8cm

                     {\hfill IC/95/125}

\vspace*{-.3cm}      {\hfill TUM--HEP--221/95}

\vspace*{-.3cm}      {\hfill MPI--PhT/95--35}

\vspace*{-.3cm}      {\hfill FTUV/95-34, IFIC/95-36}

\vspace*{-.3cm}      {\hfill July 1995}

  \vspace*{1.0cm}
  {\begin{center}       {\Large\bf
                        Left--Right~Symmetry~Breaking in NJL Approach}
                        \end{center}  }
  \vspace*{1.cm}
  {\begin{center}       {\large
                        {\sc Eugeni Akhmedov\footnote{On leave from National
                        Research Center ``Kurchatov Institute'', 123182 Moscow,
                        Russia}\footnote{\makebox[1.cm]{Email:}
                                  akhmedov@tsmi19.sissa.it},
                        Manfred Lindner\footnote{\makebox[1.cm]{Email:}
                                 Manfred.Lindner@Physik.TU-Muenchen.DE},
                        Erhard Schnapka\footnote{\makebox[1.cm]{Email:}
                                 Erhard.Schnapka@Physik.TU-Muenchen.DE}}\\
                        and
                        {\sc Jos\'e W. F.Valle\footnote{\makebox[1.cm]{Email:}
                                 valle@flamenco.ific.uv.es}}
                        }\end{center} }
   \vspace*{-0.2cm}
  {\it \begin{center}  \footnotemark[2]International Centre for
                       Theoretical Physics,\\ Strada Costiera 11,
                       I--34100 Trieste, ITALY

                       \vskip .1cm

                       \footnotemark[3]$\ \!\!^,$\footnotemark[4]
                       Institut f\"ur Theoretische Physik,
                       Technische Universit\"at M\"unchen,          \\
                       James--Franck--Strasse, D--85748 Garching, GERMANY

                       \vskip .1cm

                       \footnotemark[5]Instituto de F\'{\i}sica
                       Corpuscular - C.S.I.C.\\
                       Departament de F\'{\i}sica Te\`orica,
                       Universitat de Val\`encia\\
                       46100 Burjassot, Val\`encia, SPAIN

       \end{center} }
  \vspace*{-0.1cm}
{\Large \bf \begin{center} Abstract  \end{center}  }
We study left--right symmetric models which contain only fermion and
gauge boson fields and no elementary scalars. The Higgs bosons are generated
dynamically through a set of gauge-- and parity--invariant 4-fermion
operators. It is shown that in a model with a composite bi-doublet and two
triplet scalars there is no parity breaking at low energies, whereas in the
model with  two doublets instead of two triplets parity is broken
automatically regardless of the choice  of the parameters of the model.
For phenomenologically allowed values of the right--handed scale a tumbling
symmetry breaking mechanism is realized in which parity breaking at a high
scale $\mu_R$ propagates down and eventually causes the \EW symmetry breaking
at the scale $\mu_{EW}\sim 100~GeV$. The model exhibits a number of low and
intermediate mass Higgs bosons with certain relations between their masses.
In particular, the components of the $SU(2)_L$ Higgs doublet $\chi_L$ are
pseudo--\GBs of an accidental (approximate) $SU(4)$ symmetry of the Higgs
potential and therefore are expected to be relatively light.

\renewcommand{\baselinestretch}{1.2}

\end{titlepage}

\newpage
\renewcommand{\thefootnote}{\arabic{footnote}}
\setcounter{footnote}{0}

A few years ago a very interesting approach to \EW symmetry breaking was
put forward, the so called ``top condensate'' model \cite{N,Mir,BHL,Mar}.
In this model the low--energy degrees of freedom are just the usual fermions
and gauge bosons, i.e. no fundamental Higgs boson is present. Instead,
it is assumed that there is a strong attractive interaction in the quark
sector which can lead to the formation of  a $t\bar{t}$ bound state playing
the role of the Higgs scalar. This interaction is assumed to result from
new physics at some high--energy scale $\Lambda$, the origin and precise
nature of which is not specified. At low energies this new physics would
manifest itself through non-renormalizable interactions between the usual
fermions and gauge bosons. At energies $E \ll \Lambda$ the lowest
dimensional operators are most important, which are just the four--fermion
(4-f) operators. Assuming that the heaviest top quark drives the symmetry
breaking, one arrives at the practically unique gauge--invariant 4-f operator
\cite{N,Mir,BHL,Mar}
\beq
{\cal L}_{4f}=G(\bar{Q}_{Li}t_R)(\bar{t}_R Q_{Li})~,
\label{4f1}
\eeq
where $Q_L$ is the left--handed doublet of the third generation quarks, $G$
is a dimensionful coupling constant, $G \sim \Lambda^{-2}$, and it is
implied that the colour indices are summed over within each bracket.

The four-fermion interaction of eq.~(\ref{4f1}) can be studied analytically
in the large $N_c$ (number of colours) limit in the so--called NJL or fermion
bubble approximation\footnote{We use the well known abbreviation NJL though
the paper of Vaks and Larkin was received and published first.} \cite{VL,NJL}.
For $G>G_{\mbox{critical}}=8\pi^2/N_c\Lambda^2$ the \EW symmetry is
spontaneously broken, the top quark and the $W^\pm$ and $Z^0$ bosons acquire
masses, and a composite Higgs scalar doublet $H\sim \bar{t}_R Q_L$ is formed.
To obtain phenomenologically acceptable values for the top quark mass $m_t$
one has to assume that the 4-f coupling constant $G$ is very close to its
critical value. It has been shown \cite{BHL} that this is equivalent to the
usual fine--tuning of the Higgs boson mass in the \SMp Thus, the gauge
hierarchy problem has not been solved in the top--condensate
approach\footnote{It has been claimed in \cite{Bl} that taking into account
the loops with composite Higgs scalars results in the automatic cancelation
of quadratic divergences and solves the gauge hierarchy problem of the \SM
in the BHL approach. We do not discuss this possibility here.}. In the fermion
bubble approximation one obtains a prediction for $m_t$ which depends
logarithmically on the scale of new physics $\Lambda$ and, in addition, one
gets the relation $m_H = 2 m_t$ for the Higgs boson mass. For $\Lambda \approx
10^{15}~GeV$ one finds a value of $m_t \approx 165~GeV$. However, the \RG
improved calculations taking into account the loops with propagating
composite Higgs scalar and gauge bosons result in significantly higher
values of the top quark mass, $m_t = 220-240~GeV$ \cite{BHL}.

Nevertheless, the top condensate approach reproduces correctly the structure
of the low--energy effective Lagrangian of the \SM and demonstrates how the \EW
symmetry breaking can result from some high--energy dynamics. It is therefore
interesting to study whether a similar approach can work in various extensions
of the minimal \SMp

In this paper we consider dynamical symmetry breaking in left--right symmetric
(LR) models based on the gauge group $SU(2)_L \times SU(2)_R \times U(1)_{B-L}$
\cite{LR1,LR3}, following the BHL approach to the \SMp Left-right-symmetric
models in general are very attractive since they treat left--handed and
right--handed fermions symmetrically and explain the parity non-conservation at
low energies as a result of spontaneous symmetry breaking. It is usually
assumed that symmetry breaking in LR models occurs in two steps: first,
$SU(2)_R\times U(1)_{B-L}$ breaks down to $U(1)_Y$ at an energy scale $\mu_R$,
and second, the remaining \SM gauge group is broken down to $U(1)_{em}$ at the
\EW scale $\mu_{EW}\sim 100~GeV$. Obviously this more complicated symmetry
breaking pattern requires a richer Higgs sector, and it is interesting to
investigate whether the above symmetry breaking scenario can be successfully
reproduced in a dynamical model with composite Higgs bosons. As in the BHL
approach, we will only consider the usual fermions and gauge bosons of the
model as elementary particles, with no fundamental Higgs scalars being present,
and in addition introduce a set of relevant 4-f interactions stemming from
unspecified new physics at a high energy scale $\Lambda$. Here we derive
our conclusions in the bubble approximation; more complete results including
the \RG improved predictions will be reported elsewhere \cite{ALSV2}.

The Higgs sector of the most popular LR model \cite{LR3} consists of a
bi-doublet $\phi \sim (2,2,0)$ and two triplets, $\Delta_L \sim (3,1,2)$
and $\Delta_R \sim (1,3,2)$, where the quantum numbers with respect to the
LR gauge group are shown. Assuming that these scalars are bound states of the
usual fermions, the following fermionic content reproduces the correct quantum
numbers:
$$\phi_{ij} \sim \alpha (\bar{Q}_{Rj}Q_{Li}) + \beta (\tau_2 \bar{Q}_L Q_R
\tau_2)_{ij}+\mbox{leptonic terms}~, $$
\beq
\vec{\Delta}_L \sim (\Psi_L^T C \tau_2 \vec{\tau}\Psi_L),\;\;
\vec{\Delta}_R\sim (\Psi_R^T C  \tau_2 \vec{\tau}\Psi_R)~.
\label{phiij}
\eeq
Here $Q_L,\Psi_L$ ($Q_R,\Psi_R$) are left--handed (right--handed) doublets
of quarks and leptons, respectively; $i$ and $j$ are isospin indices.

In models with Higgs bosons generated by 4-f operators the composite scalars
are, roughly speaking, ``square roots'' of these 4-f operators. One can
therefore obtain the above composite Higgs bosons starting from the 4-f
operators which are ``squares'' of the expressions in eq.~(\ref{phiij}).
A convenient way to study models with composite Higgs bosons is the
auxiliary field technique, in which one introduces the static auxiliary
scalar fields (with appropriate quantum numbers) with Yukawa couplings and
mass terms but no kinetic terms and no quartic couplings. Since the modified
Lagrangian of the system is quadratic in these auxiliary fields they can always
be integrated out in the functional integral \cite{Auxform}. Equivalently,
one can use the equations of motion for these fields to express them in
terms of the fermionic degrees of freedom. After substituting the resulting
expressions into the auxiliary Lagrangian one reproduces the initial 4-f
structures.

The static auxiliary fields can acquire gauge--invariant kinetic terms and
quartic self--interactions through radiative corrections and become
physical propagating scalar fields at low energies provided that the
corresponding gap equations are satisfied  \cite{BHL}. The kinetic terms and
mass corrections can be derived from the 2--point Green function, whereas
the quartic couplings are given by the 4--point functions. Given the Yukawa
couplings of the scalar fields one can readily calculate these functions in
the fermion bubble approximation, in which they are given by the corresponding
1-fermion--loop diagrams.

Consider now spontaneous parity breakdown in LR models with composite Higgs
bo\-sons. It is usually assumed that, in addition to the gauge symmetry, the
Lagrangian of the LR model possesses the discrete parity symmetry under which
\beq
Q_L \leftrightarrow Q_R,\;\;\; \Psi_L \leftrightarrow
\Psi_R,\;\;\; \phi \leftrightarrow \phi^\dagger,\;\;\;
\Delta_L \leftrightarrow \Delta_R,\;\;\;W_L\leftrightarrow W_R~.
\label{discrete}
\eeq
Even if the Higgs potential of the model is exactly symmetric with respect
to the discrete parity transformation, parity can be spontaneously broken
by $\langle \Delta_R \rangle >  \langle \Delta_L \rangle$ \cite{LR2}.
It is easily seen that this can only occur provided $\lambda_2 >
\lambda_1$ where $\lambda_1$ and $\lambda_2$ are the coefficients of the
$[(\Delta_L^\dagger \Delta_L)^2+(\Delta_R^\dagger \Delta_R)^2]$ and
$2(\Delta_L^\dagger \Delta_L)(\Delta_R^\dagger \Delta_R)$ quartic couplings
in the Higgs potential. While in the conventional approach $\lambda_1$ and
$\lambda_2$ can be chosen appropriately as free parameters of the model, the
scalar mass terms and couplings in the composite Higgs approach are not
arbitrary; they are all calculable in terms of the 4-f couplings $G_a$ and
the scale of new physics $\Lambda$ \cite{BHL}. In particular, in the fermion
bubble approximation at one loop level the quartic couplings $\lambda_1$ and
$\lambda_2$ are induced through the Majorana--like Yukawa couplings
$f(\Psi_L^T C \tau_2 \vec{\tau}\vec{\Delta}_L \Psi_L+\Psi_R^T C\tau_2
\vec{\tau}\vec{\Delta}_R \Psi_R)+h.c.$, and are given by the diagrams of
Fig.~1.
\begin{figure}[htb]
\centerline{
\rotate[r]{\epsfxsize=31ex \epsffile{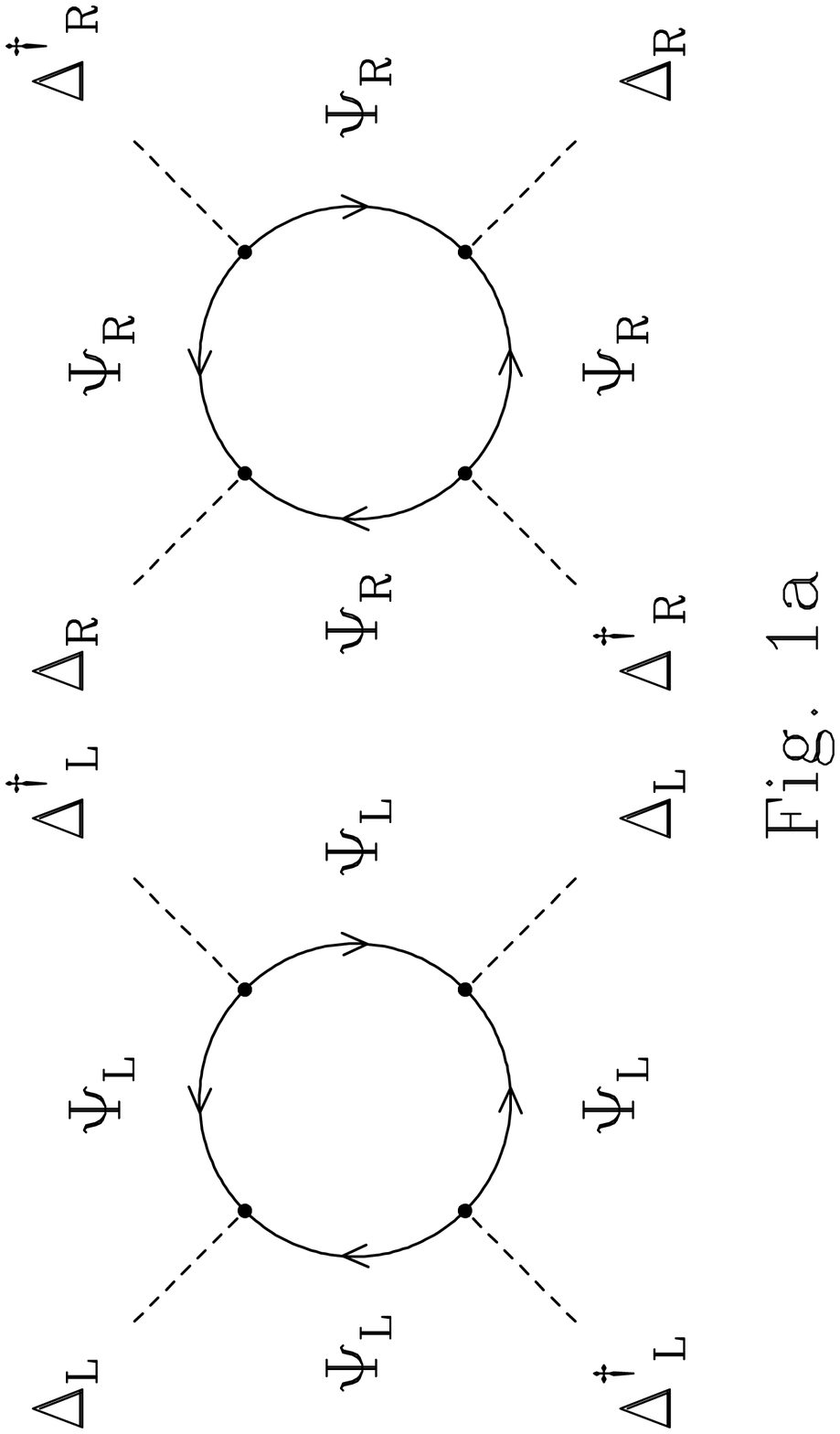} }
\hspace{2em}
\rotate[r]{\epsfxsize=31ex \epsffile{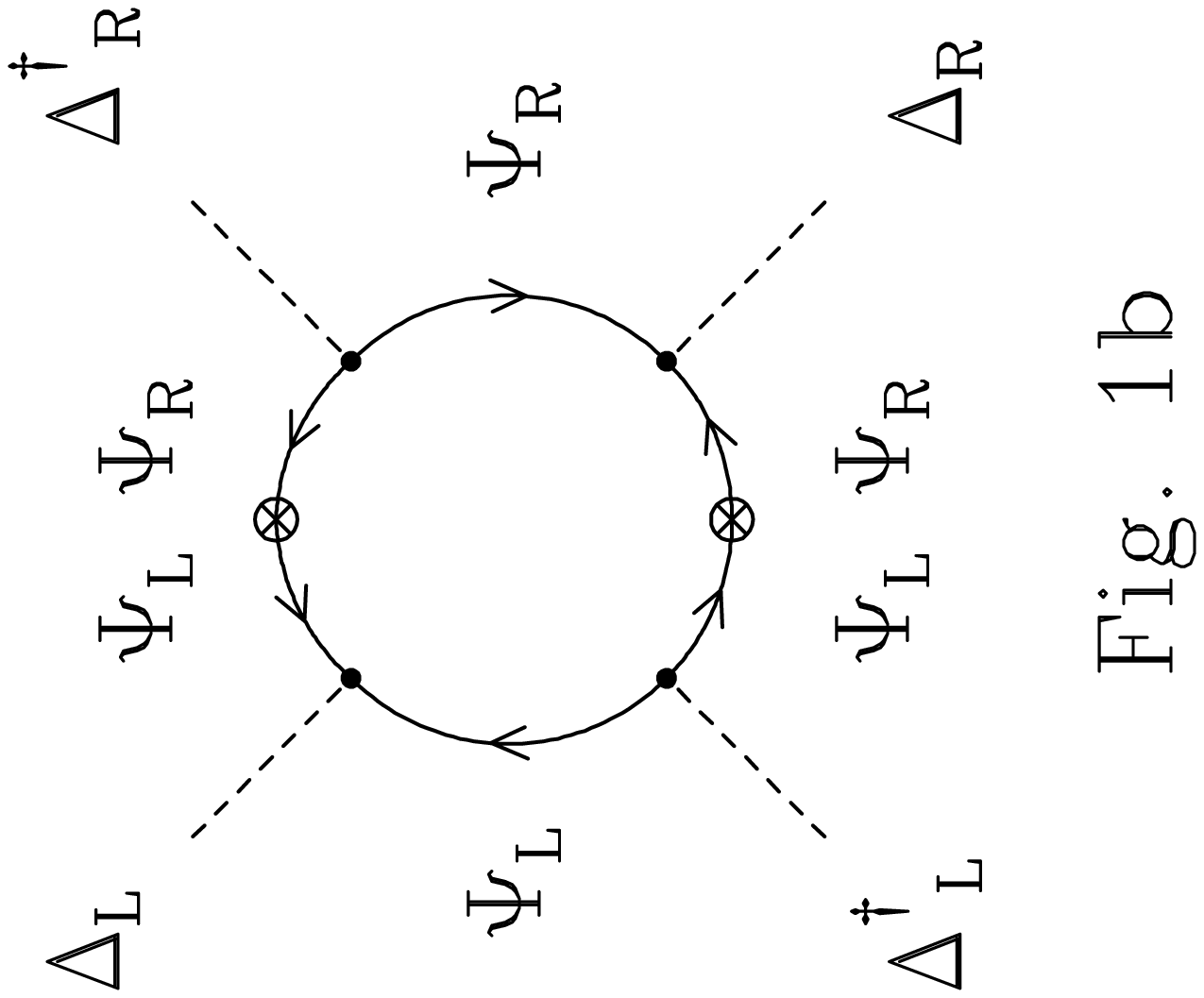} }
}
\caption[]{\small\sl
Fermion loop diagrams contributing to the quartic couplings
$\lambda_1$~(Fig.~\ref{fig:delta}a) and $\lambda_2$~(Fig.~\ref{fig:delta}b)
for Higgs triplets.
\label{fig:delta}
}
\end{figure}
It can be seen from Fig.~1b that to induce the $\lambda_2$ term one needs the
$\Psi_L$--$\Psi_R$ mixing in the fermion line in the loop, i.e. the lepton
Dirac mass term insertions. However, the Dirac mass terms are generated by
the VEVs of the bi-doublet $\phi$; they are absent at the parity breaking
scale which is supposed to be higher than the \EW scale. Even if parity and
\EW symmetry are broken simultaneously (which is hardly a phenomenologically
viable scenario), this would not save the situation since the diagram of
Fig.~1b is finite in the limit $\Lambda\to\infty$ whereas the one of Fig.~1a
is logarithmically divergent and so the inequality $\lambda_2>\lambda_1$ cannot
be satisfied.

Therefore one is lead to consider a model with a different composite
Higgs content. The simplest LR model \cite{LR1} includes two doublets,
$\chi_L \sim (2,1,-1)$ and $\chi_R \sim (1,2,-1)$, instead of the triplets
$\Delta_L$ and $\Delta_R$. As we shall see, the model with composite doublets
will automatically result in the correct pattern of the dynamical breakdown
of parity.

Since we want the doublet scalars to be composite, we require an additional
gauge--singlet fermion as a necessary constituent. We therefore assume that in
addition to the usual quark and lepton doublets there is a gauge--singlet
fermion
\beq
S_L \sim (1,\;1,\;0)~.
\label{SL}
\eeq
To maintain the discrete parity symmetry one needs a right--handed counterpart
of $S_L$. This can be either another particle, $S_R$, or the right--handed
antiparticle of $S_L$, $(S_L)^c\equiv C\bar{S}_L^T = S^c_R$. The latter choice
is more economical and, as we shall see, leads to the desired symmetry
breaking pattern. We therefore assume that under parity operation
\beq
S_L \leftrightarrow S^c_R~.
\label{SLC}
\eeq
With this new singlet and the usual quark and lepton doublets we introduce
the following set of gauge--invariant 4-f interactions:
\bea
{\cal L}_{4f}=G_1(\bar{Q}_{Li}Q_{Rj})(\bar{Q}_{Rj}Q_{Li})+
[G_2(\bar{Q}_{Li}Q_{Rj})(\tau_2\bar{Q}_{L}Q_{R}\tau_2)_{ij}
+h.c.]\nonumber \\
+G_3(\bar{\Psi}_{Li}\Psi_{Rj})(\bar{\Psi}_{Rj}\Psi_{Li})+
[G_4(\bar{\Psi}_{Li}\Psi_{Rj})(\tau_2\bar{\Psi}_{L}\Psi_{R}\tau_2)_{ij}
+h.c.]\nonumber \\
+[G_5(\bar{Q}_{Li}Q_{Rj})(\bar{\Psi}_{Rj}\Psi_{Li})+h.c.]+
[G_6(\bar{Q}_{Li}Q_{Rj})(\tau_2\bar{\Psi}_{L}\Psi_{R}\tau_2)_{ij}
+h.c.] \nonumber \\
+G_7[(S_L^T C \Psi_L)(\bar{\Psi}_L C \bar{S}_L^T)+
(\bar{S}_L\Psi_R)(\bar{\Psi}_R S_L)]+G_8 (S_L^T C S_L)(\bar{S}_L C
\bar{S}_L^T)~.
\label{L4f}
\eea
In analogy to the BHL model the $G_a$ are dimensionful 4-f couplings
of the order of $\Lambda^{-2}$ motivated by some new physics at $\Lambda$.
Notice that the above interactions are not only gauge--invariant, but also
(for hermitian $G_2$, $G_4$, $G_5$ and $G_6$) symmetric with respect to the
discrete parity operation (\ref{discrete}), (\ref{SLC}).

We assume that only the third generation fermions contribute to
${\cal L}_{4f}$, i.e., deal with a limit where only the heaviest fermions
are massive while all the light fermions are considered to be massless.
This seems to be a good starting point from where light fermion masses could,
e.g.,  be generated radiatively. In addition to the bidoublet $\phi$
of the structure given in eq.~(\ref{phiij}), the above 4-f couplings, if
critical, can give rise to a pair of composite doublets $\chi_L$ and $\chi_R$
and also to a singlet scalar $\sigma$:
\beq
\chi_L \sim S_L^T C \Psi_L, \;\;\;\;  \chi_R \sim \bar{S}_L\Psi_R =
(S^c_R)^T C \Psi_R, \;\;\;\; \sigma \sim \bar{S}_L C \bar{S}_L^T~.
\label{composite}
\eeq
{}From eqs.~(\ref{discrete}) and (\ref{SLC}) it follows that under parity
we have $\chi_L \leftrightarrow \chi_R$ and $\sigma \leftrightarrow
\sigma^\dagger$. Switching to the auxiliary field formalism, the scalars
$\chi_L$, $\chi_R$, $\phi$ and $\sigma$ have the following bare mass terms
and Yukawa couplings:
\bea
L_{aux}&=&-M_0^2(\chi_L^\dagger \chi_L+\chi_R^\dagger \chi_R)-M_1^2
\tr{(\phi^\dagger \phi)}-\frac{M_2^2}{2}\tr{(\phi^\dagger
\tilde{\phi}+h.c.)}-M_3^2 \sigma^\dagger\sigma \nonumber \\
& & -\left[\bar{Q}_L(Y_1\phi+Y_2\tilde{\phi})Q_R +
\bar{\Psi}_L(Y_3\phi+Y_4\tilde{\phi})\Psi_R + h.c.\right] \nonumber \\
& &-\left[Y_5(\bar{\Psi}_L \chi_L S^c_R+\bar{\Psi}_R \chi_R S_L)
+Y_6 (S_L^T C S_L)\sigma + h.c.\right]~,
\label{Laux}
\eea
where the field $\tilde{\phi}\equiv \tau_2\phi^*\tau_2$ has the same quantum
numbers as $\phi$: $\tilde{\phi}\sim (2,\;2,\;0)$. By integrating out the
auxiliary scalar fields one can reproduce the 4-f structures of
eqs.~(\ref{L4f}) and express the 4-f couplings $G_1,...,G_8$ in terms of the
Yukawa couplings $Y_1,...,Y_6$ and the mass parameters $M_0^2$, $M_1^2$,
$M_2^2$ and $M_3^2$ (explicit formulas can be found in \cite{ALSV2}).
In components, the scalar multiplets of the model are
\beq
\phi =  \left( \begin{array}{cc}
\phi_1^0 & \phi_2^+ \\
\phi_1^-& \phi_2^0 \end{array} \right)\,,\;\;\;
\langle \phi \rangle = \left( \begin{array}{cc}
\kappa & 0 \\
0 & \kappa' \end{array} \right)\,,\;\;\;
\chi_{L}=\left( \begin{array}{c}
\chi_L^0 \\ \chi_L^-\end{array} \right) \,,\;\;\;
\chi_{R}=\left( \begin{array}{c}
\chi_R^0 \\ \chi_R^-\end{array} \right) ~.
\label{bidef}
\eeq

Let us now consider parity breaking in the present LR model. In a viable
scenario the $SU(2)_R$ symmetry should be broken at the right--handed scale
$\mu_R$ by $\langle \chi_R^0 \rangle = v_R$, and the \EW symmetry has to be
broken at $\mu_{EW}$ by the VEVs of $\phi$ and possibly of $\chi_L^0$
($\equiv v_L$). Using the Yukawa couplings of the doublets $\chi_L$ and
$\chi_R$ (see eq.~(\ref{Laux})), one can calculate the fermion-loop
contributions to the  quartic couplings $\lambda_1 [(\chi_L^\dagger\chi_L)^2
+(\chi_R^\dagger \chi_R)^2]$ and $2\lambda_2 (\chi_L^\dagger\chi_L)
(\chi_R^\dagger \chi_R)$ in the effective Higgs potential (Fig.~2a and 2b).
\begin{figure}[htb]
\centerline{
\rotate[r]{\epsfxsize=31ex \epsffile{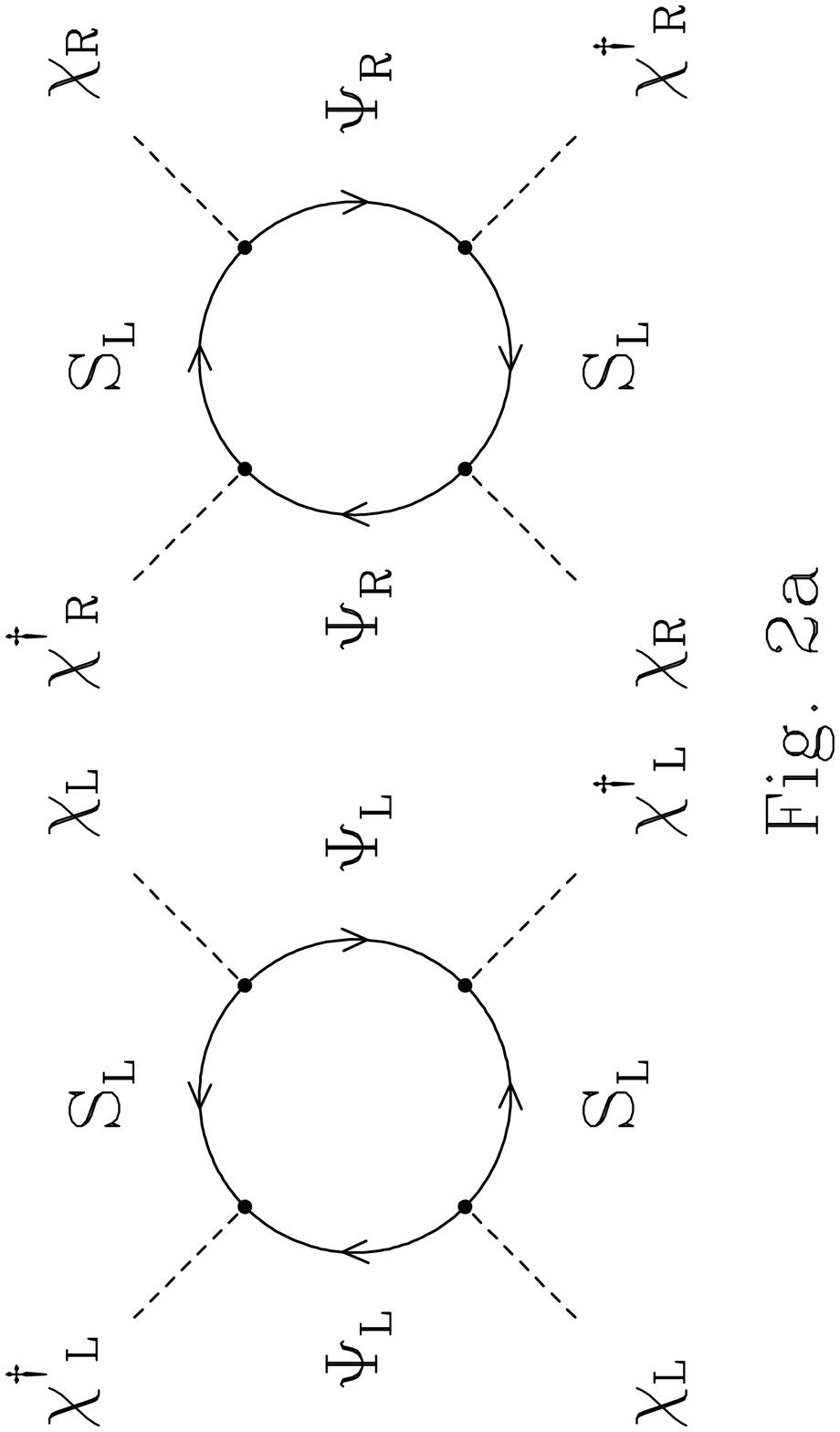} }
\hspace{2em}
\rotate[r]{\epsfxsize=31ex \epsffile{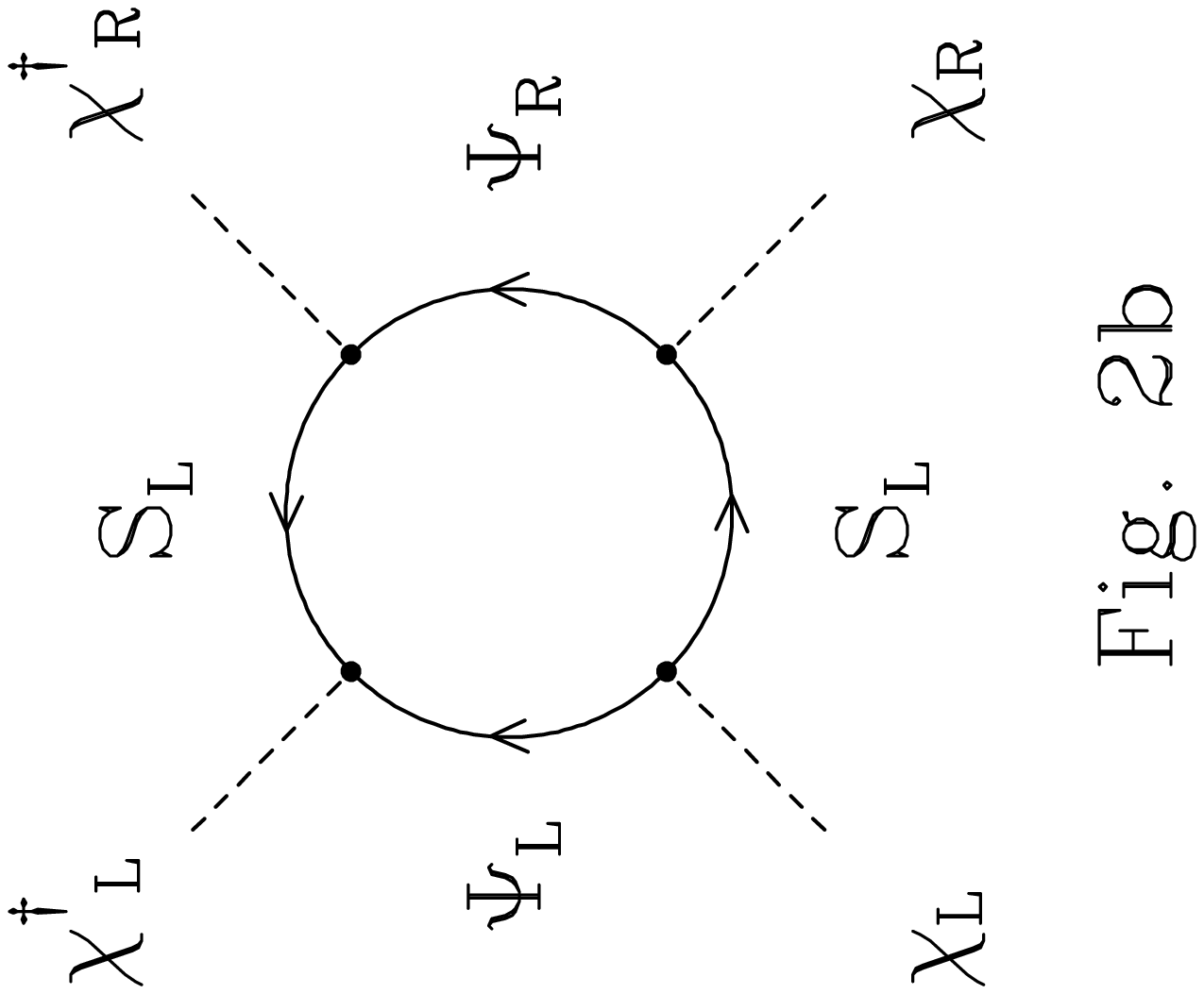} }
}
\caption[]{\small\sl
Fermion loop diagrams contributing to the quartic couplings
$\lambda_1$~(Fig.~\ref{fig:doub}a) and
$\lambda_2$~(Fig.~\ref{fig:doub}b) for the Higgs doublets $\chi_{L/R}$.
\label{fig:doub}
}
\end{figure}
The $\lambda_1$ and $\lambda_2$ terms are now given by similar diagrams.
Since the Yukawa couplings of $\chi_L$ and $\chi_R$ coincide (which is just
the consequence of the discrete parity symmetry), Figs.~2a and 2b yield
$\lambda_1=\lambda_2$. Recall that one needs $\lambda_2> \lambda_1$
to have spontaneous parity breakdown in the LR models. As we shall see,
taking into account the gauge boson loop contributions to $\lambda_1$ and
$\lambda_2$ will automatically secure this relation.

Both $\lambda_1$ and $\lambda_2$ obtain corrections from $U(1)_{B-L}$ gauge
boson loops, whereas only $\lambda_1$ is corrected by diagrams with $W^i_L$
or $W^i_R$ loops (see Fig.~3).
\begin{figure}[htb]
\centerline{
\rotate[r]{\epsfxsize=31ex \epsffile{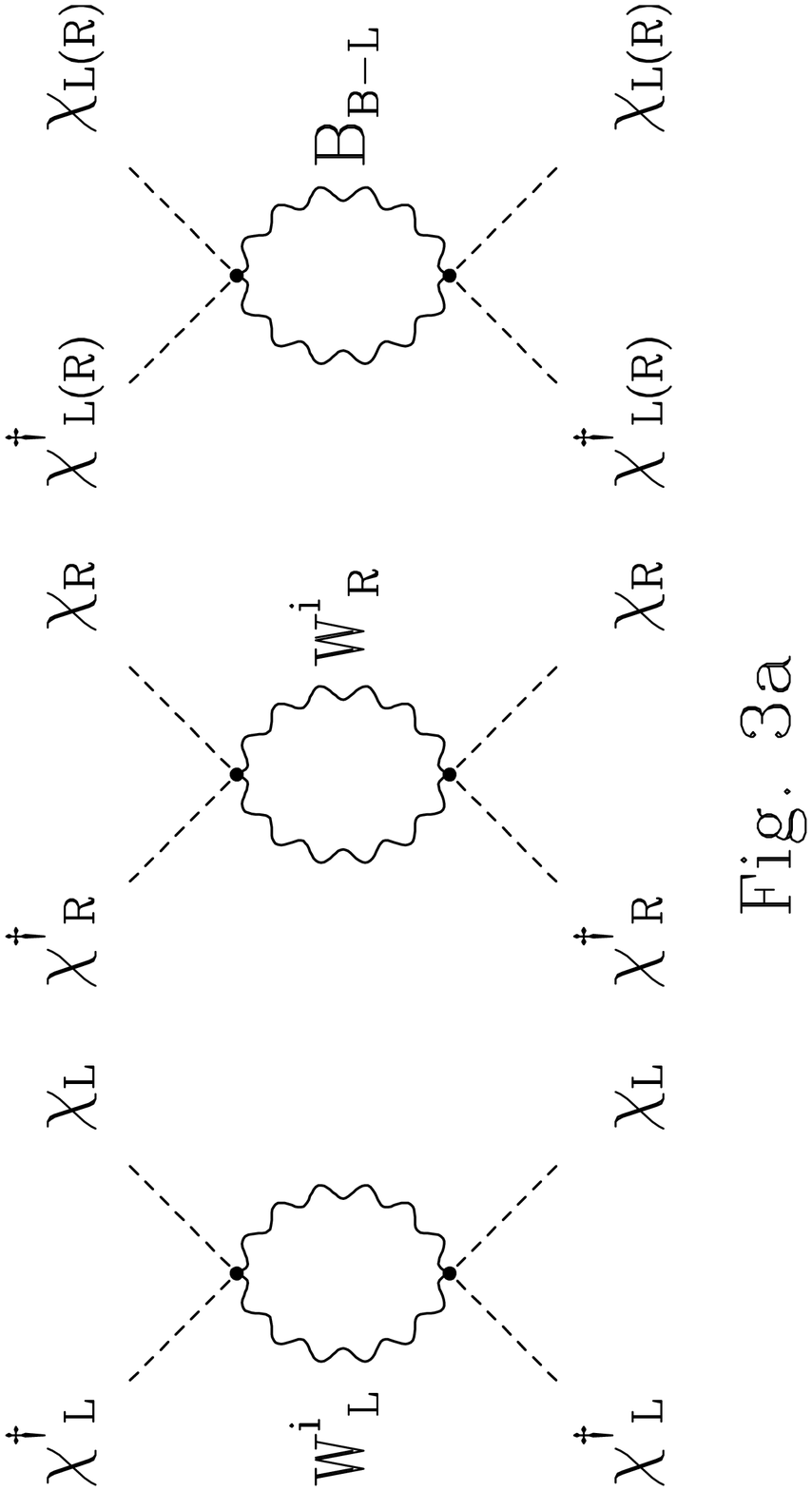} }
\hspace{2em}
\rotate[r]{\epsfxsize=31ex \epsffile{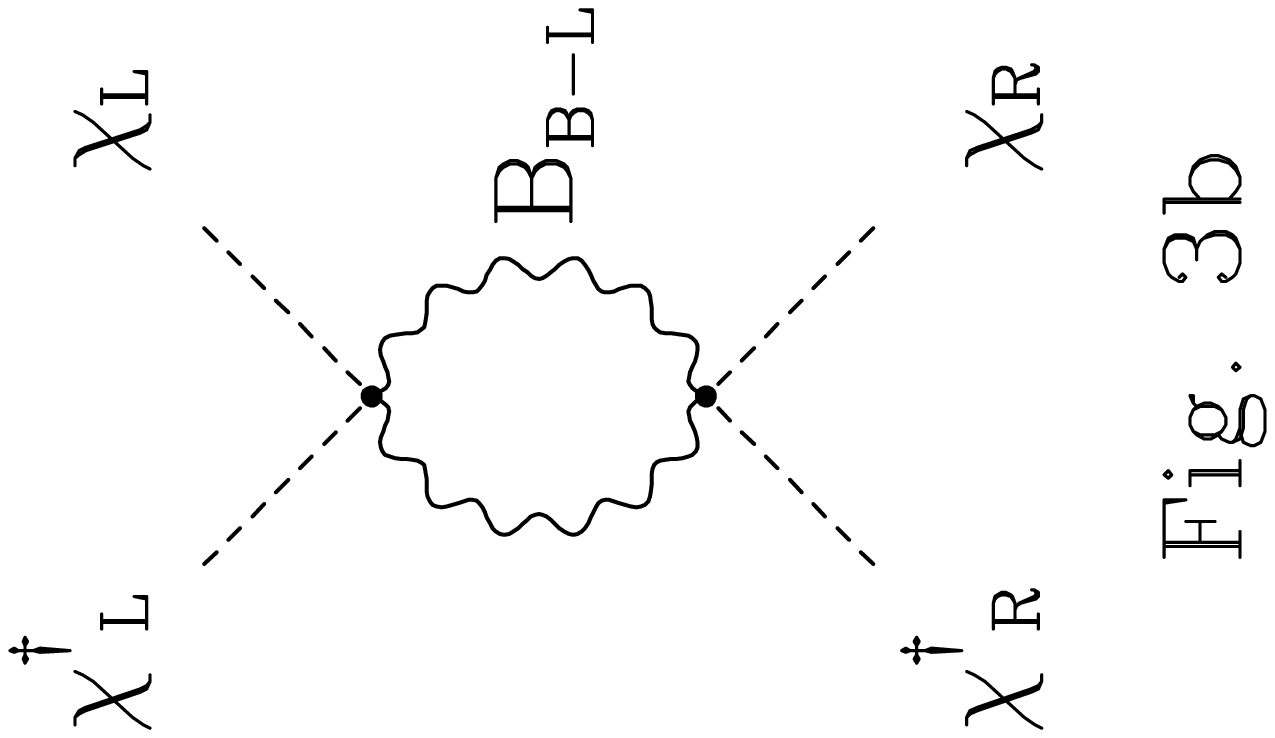} }
}
\caption[]{\small\sl
Gauge boson loop diagrams contributing to the quartic couplings
$\lambda_1$~(Fig.~\ref{fig:gauge}a) and
$\lambda_2$~(Fig.~\ref{fig:gauge}b) for the Higgs doublets $\chi_{L/R}$
in Landau gauge.
\label{fig:gauge}
}
\end{figure}
Since all these contributions have a relative minus sign compared to the
fermion loop ones, one finds $\lambda_2>\lambda_1$ irrespective of the
values of the Yukawa or gauge couplings or any other parameter of the model,
provided that the $SU(2)$ gauge coupling $g_2\neq 0$ [compare the expressions
for $\lambda_1$ and $\lambda_2$ in~(\ref{Lambda}) below]. Thus the condition
for spontaneous parity breakdown is automatically satisfied in our model.

We have a very interesting situation here. In a model with composite triplets
$\Delta_L$ and $\Delta_R$ parity is never broken, i.e. the model is not
phenomenologically viable. At the same time, in the model with two composite
doublets $\chi_L$ and $\chi_R$ instead of two triplets (which requires
introduction of an additional singlet fermion $S_L$) parity is broken
automatically. This means that, unlike in conventional LR models, in the
composite Higgs approach {\em whether or not parity can be spontaneously
broken depends on the particle content of the model rather than on the
choice of the parameters of the Higgs potential}.

{}From eq.~(\ref{Laux}) one can readily find the fermion masses. The masses
of the quarks and charged leptons and the Dirac neutrino mass $m_D$ are
given by the VEVs of the bi--doublet (we assume all the VEVs to be real):
\bea
m_t=Y_1\kappa+Y_2\kappa',\;\;\;\;\;m_D=Y_3\kappa+Y_4\kappa',\nonumber \\
m_b=Y_1\kappa'+Y_2\kappa,\;\;\;\;\; m_{\tau}=Y_3\kappa'+Y_4\kappa.
\label{fmass}
\eea
It is well known that LR models with only doublet Higgs scalars usually
suffer from the large neutrino mass problem. It turns out that introducing
the singlet fermion $S_L$ not only provides the spontaneous parity breaking
in our model, but also cures the neutrino mass problem. In fact, as it was
first noticed in \cite{WW}, with an additional singlet neutral fermion $S_L$
the neutrino mass matrix takes the form (in the basis $(\nu_L, \nu^c_L, S_L)$)
\beq
M_{\nu} = \left(
\begin{array}{c}
\end{array}
\begin{array}{ccc}
0 & m_D & \beta\\
m_D & 0 & M\\
\beta & M & \tilde{\mu}
\end{array} \right)\;,
\label{numass1}
\eeq
where the entries $\beta$, $M$ and $\tilde{\mu}$ can be read off from
eq.~(\ref{Laux}),
\beq
\beta = Y_5 v_L \, , \quad M = Y_5 v_R \, \quad  \tilde{\mu} = 2Y_6\sigma_0~,
\label{yukawas}
\eeq
with $\sigma_0\equiv \langle\sigma\rangle$. For $v_R \gg \kappa, \kappa',
v_L$ and $v_R \gta \sigma_0$ one obtains two heavy Majorana
neutrino mass eigenstates  with the masses $\sim M$ and a light Majorana
neutrino with the mass $m_\nu \simeq \tilde{\mu}(m_D^2/M^2)-2\beta m_D/M$
which vanishes in the limit $M \to \infty$. This is the modified seesaw
mechanism which provides the smallness of neutrino mass.

As we mentioned before, radiative corrections in the auxiliary field
formalism result in gauge--invariant kinetic terms, quartic interactions
and renormalized mass terms for the scalar fields at low energies
$E < \Lambda$. The effective low--energy Lagrangian of the system in
the bubble approximation can be written as\footnote{For a detailed
derivation of ${\cal L}_{\mbox{eff}}$ see \cite{ALSV2}.}
\bea
{\cal L}_{\mbox{eff}}={\cal L}_0+Z_\phi \tr \[(D^\mu \phi)^\dagger
(D_\mu \phi)\]+Z_\chi\[(D^\mu \chi_L)^\dagger (D_\mu \chi_L)+
(D^\mu \chi_R)^\dagger (D_\mu \chi_R)\]  \nonumber \\
+Z_\sigma (\partial^\mu \sigma)^\dagger (\partial_\mu \sigma)
+{\cal L}_{Yuk}+V_{\mbox{eff}}~,
\label{Leff}
\eea
where ${\cal L}_0$ contains the gauge--invariant kinetic terms of fermions
and gauge bosons and  ${\cal L}_{Yuk}$ is given by the Yukawa--coupling
terms in eq.~(\ref{Laux}). The scalar wave-function renormalization
constants are
\bea
Z_{\phi} & = & \sixtpi\, \[N_c (Y_{1}^{2}+Y_{2}^{2})+Y_{3}^{2}+Y_{4}^{2}\]\,
\lnlambda\,, \nonumber \\
Z_{\chi} & = & \sixtpi\, Y_{5}^{2}\, \lnlambda \, , \quad
Z_{\sigma}\; =\;  \sixtpi\, 2 Y_6^2\, \lnlambda~.
\label{zfactors}
\eea
Further, the effective Higgs potential $V_{\mbox{eff}}$
in eq.~(\ref{Leff}) is given by
\bea
V_{\mbox{eff}}&= &\tilde{M}_0^2(\chi_L^\dagger\chi_L+\chi_R^\dagger\chi_R)
          +\tilde{M}_1^2 \tr{(\phi^\dagger \phi)}+\frac{\tilde{M}_2^2}{2}
           \tr{(\phi^\dagger \tilde{\phi}+h.c.)}+\tilde{M}_3^2
           \sigma^\dagger\sigma \nonumber \\
& & +\lambda_1[(\chi_L^\dagger \chi_L)^2+(\chi_R^\dagger \chi_R)^2]
    +2\lambda_2 (\chi_L^\dagger \chi_L)(\chi_R^\dagger \chi_R)
     + \frac{1}{2}\lambda_3[\chi_L^\dagger(Y_3\phi+Y_4 \tilde{\phi})\chi_R
       \sigma^\dagger + h.c.] \nonumber \\
& &  +\lambda_4[\chi_L^\dagger(Y_3\phi+Y_4 \tilde{\phi})
                             (Y_3\phi^\dagger+Y_4\tilde{\phi}^\dagger)\chi_L
              +\chi_R^\dagger(Y_3\phi^\dagger+Y_4\tilde{\phi}^\dagger)
                             (Y_3\phi+Y_4 \tilde{\phi})\chi_R] \nonumber \\
& & +\lambda_5(\chi_L^\dagger \chi_L+\chi_R^\dagger \chi_R)\tr(\phi^\dagger
   \phi)
    +\lambda_6(\chi_L^\dagger\chi_L+\chi_R^\dagger \chi_R)\sigma^\dagger
      \sigma\nonumber \\
& & +\lambda_7' \tr(\phi^\dagger\phi\phi^\dagger\phi)
    +\frac{1}{3}\lambda_8'\tr(\phi^\dagger\tilde{\phi}
      \tilde{\phi}^\dagger\phi)
    +\frac{1}{12}\lambda_8'
        [\tr(\phi^\dagger\tilde{\phi}\phi^\dagger\tilde{\phi})+h.c.]
  \nonumber \\
& &  +\frac{1}{2} \lambda_9
        [\tr(\phi^\dagger\phi\phi^\dagger\tilde{\phi})+h.c.]
    +\lambda_0
        [\tr(\phi^\dagger\phi)]^2 + \lambda_{10}(\sigma^\dagger\sigma)^2~.
\label{Veff}
\eea
Here we give explicitly only the mass terms and the quartic couplings which
we will refer to later, the complete set is given in \cite{ALSV2}.
\bea
\tilde{M}_0^2 & = & M_0^2-\frac{1}{8\pi^2}\left [Y_5^2-\frac{3}{8}
Z_{\chi}(3g_2^2+g_1^2)\right ] (\Lambda^2-\mu^2)
\label{M02} \\
\tilde{M}_1^2 & = & M_0^2-\frac{1}{8\pi^2}\left \{ \left [N_c(Y_1^2+Y_2^2)+
(Y_3^2+Y_4^2)\right ]-\frac{9}{4}Z_{\phi}g_2^2\right \}(\Lambda^2-\mu^2)
\label{M12}  \\
\tilde{M}_2^2 & = & M_2^2-\frac{1}{4\pi^2}(N_c Y_1 Y_2+Y_3 Y_4)
(\Lambda^2-\mu^2)
\label{M22} \\
\tilde{M}_3^2 & = & M_0^2-\frac{1}{4\pi^2}\,Y_6^2 \, (\Lambda^2-\mu^2)
\label{M32} \\
\lambda_1 & = & \sixtpi\[ Y_5^4-\frac{3}{16}(3g_2^4+2g_2^2g_1^2+
g_1^4)Z_\chi^2\]
\lnlambda \, \nonumber \\
\lambda_2 & = & \sixtpi\[ Y_5^4-\frac{3}{16}g_1^4 Z_\chi^2 \] \lnlambda
 \nonumber \, \\
\lambda_0 & = & \sixtpi\[-\frac{3}{2}g_2^4 Z_\phi^2\] \lnlambda \,,\quad\quad
  \quad\;
\lambda_5 \; = \; \sixtpi\[ -\frac{9}{8} g_2^4 Z_\phi Z_\chi \] \lnlambda \,
  \nonumber \\
\lambda_7' & =&\sixtpi\[ N_c
(Y_1^4+Y_2^4)+(Y_3^4+Y_4^4)\]\lnlambda \;, \quad \quad \quad
   \lambda_7 \; = \; \lambda_7' + \lambda_0~.
\label{Lambda}
\eea
Here $g_2$ and $g_1$ are the $SU(2)$ and $U(1)_{B-L}$ gauge couplings,
respectively. The parameters of the above effective Lagrangian
depend on the energy scale $\mu$, i.e. they are the running
parameters\footnote{This bubble--approximation running exactly coincides with
the running one would get from 1--loop \RG equations keeping only trace terms
in the relevant $\beta$ functions and imposing the compositeness boundary
conditions \cite{BHL}. The results of the \RG study with the full 1--loop
$\beta$ functions will be reported in \cite{ALSV2}.}.
At $\mu \to \Lambda$ the kinetic terms and quartic couplings of the scalar
fields vanish, their mass terms are driven towards their bare values, and one
recovers the Lagrangian with auxiliary static scalar fields.

While the bare mass parameters $M_i^2$ in eq.~(\ref{Laux}) are positive, the
corresponding running quantities $\tilde{M}_i^2$, given by eqs.~(\ref{M02})
-- (\ref{M32}), may become negative at low energy scales provided that the
corresponding Yukawa couplings are large enough. Those values for which this
occurs at $\mu=0$ we shall call the {\em critical } Yukawa couplings. For
$\tilde{M}_i^2$ to become negative at some scale $\mu^2 > 0$ the corresponding
Yukawa couplings or combinations of them must be above their critical values.
If this is to happen at scales $\mu \ll \Lambda$ the Yukawa couplings
must be fine-tuned very closely to their critical values to ensure
the proper cancelation between the large bare masses of the scalars
and the $\Lambda^2$ corrections in eqs.~(\ref{M02}) -- (\ref{M32}).
This is equivalent to the usual fine--tuning problem of gauge theories with
elementary Higgs scalars \cite{BHL}.

We assume that the scale $\mu_R$ at which parity gets spontaneously broken
(i.e. $\chi_R^0$  develops a VEV) is higher than the \EW scale
$\mu_{EW}\sim 100~GeV$, i.e. that $\tilde{M}_0^2$ changes its sign at
a higher scale than $\tilde{M}_1^2$. This means that
$Y_5^2 - (3/8)Z_\chi(3g_2^2+g_1^2)$ should be bigger than
$\tilde{Y}^2-\frac{9}{4}Z_{\phi}g_2^2$ [see eqs.~(\ref{M02}) and (\ref{M12})],
where $\tilde{Y}^2 \equiv N_c(Y_1^2+Y_2^2)+(Y_3^2+Y_4^2)\;$.
The analysis of the vacuum structure in our model \cite{ALSV2}
shows that if the condition
\beq
Y_5^2-\frac{3}{8}Z_\chi(3g_2^2+g_1^2)>2\,Y_6^2~
\label{condit}
\eeq
is satisfied, either $\chi_R$ or $\chi_L$ (but not both) acquire a VEV but
the $\sigma$ field does not, whereas for the opposite condition $\sigma$
acquires a non-zero VEV but not $\chi_R$ or $\chi_L$. Clearly the latter
situation is phenomenologically unacceptable, but by choosing the 4-f
couplings $G_7$ and $G_8$ accordingly \cite{ALSV2} we can easily
satisfy eq.~(\ref{condit}).

Let us now discuss the vacuum structure below the \EW breaking scale.
The non-vanishing VEVs are $v_R$, $\kappa$ and $\kappa'$. Since $m_t \gg
m_b$, it follows from  eq.~(\ref{fmass}) that $\kappa$ should be much larger
than $\kappa'$ or vice versa provided no significant cancelation between
$Y_1\kappa'$ and $Y_2\kappa$ occurs. Without loss of generality one can
take $\kappa \gg \kappa'$. To further simplify the discussion, we shall make
the frequently used assumption \cite{LR} $\kappa'=0$. The relation
$m_t \gg m_b$ then translates into $Y_1 \gg Y_2$. In the conventional
approach this assumption does not lead to any contradiction with
phenomenology. However, as we shall see, in our case the condition
$\kappa'=0$ cannot be exact.

Consistency of the first--derivative conditions with $\kappa'=0$ requires
$Y_1 Y_2=0$, $Y_3 Y_4=0$ and $M_2^2=0$ (this gives $\tilde{M}_2^2=\lambda_9
=0$, and as follows from eq.~(\ref{Veff}), all the terms in the effective
potential which are linear in $\kappa'$ become zero in this limit, as they
should). The condition $Y_1 Y_2=0$ along with $\kappa'=0$ implies that
either $Y_1=0$, $m_t=0$ or $Y_2=0$, $m_b=0$. The first possibility is
obviously phenomenologically unacceptable whereas the second one can be
considered as a reasonable first approximation; we therefore assume
$Y_1 \neq 0$ and $Y_2=0$. The situation is less clear for the lepton
Yukawa couplings $Y_3$ and $Y_4$. Since $m_\tau \ll m_t$ and the Dirac
mass $m_D$ of $\nu_\tau$ is unknown, one can choose either $Y_3 \neq 0$,
$Y_4=0$ or $Y_3=0$, $Y_4 \neq 0$. It turns out that the vacuum stability
condition in our model requires $Y_4^2 >Y_3^2$, therefore we choose $Y_3=0$
and $Y_4\neq 0$.

For $\sigma_0=v_L=\kappa'=Y_2=Y_3=0$ one can easily find
expressions for the VEVs of $\chi_R$ and $\phi$ \cite{ALSV2}.
Approximate expressions in terms of the parity breaking scale $\mu_R$ and
the \EW breaking scale $\mu_{EW}$ are
\beq
v_R^2 \simeq \left(\frac{M_0^2}{\Lambda^2}\right)
\frac{\mu_{R}^2}{2\lambda_1}\,, \quad\quad
\kappa^2\simeq \left( \frac{M_0^2}{\Lambda^2}\right)
\frac{\mu_{EW}^2}{2\lambda_7}~,
\label{k20}
\eeq
and the ratio of the squared VEVs can be written as
\beq
\frac{\kappa^2}{v_R^2}\simeq \left(\frac{\lambda_1}{\lambda_7}\right)
\frac{\mu_{EW}^2}{\mu_R^2}\sim \frac{\mu_{EW}^2}{\mu_R^2}\simeq
\frac{|\lambda_5|}{2\lambda_1}+\frac{\mu_1^2}{\mu_R^2}~.
\label{ratio}
\eeq

The parity breaking scale $\mu_R$ is the scale where
the effective mass term $\tilde{M}_0^2$ becomes negative for a given
Yukawa coupling $Y_5>(Y_5)_{crit}$ (formally $\mu_R^2<0$ for sub--critical
$Y_5$), while $\mu_1$ is the scale, different from $\mu_{EW}$, where this
happens for the mass term $\tilde{M}_1^2$ and a given $\tilde{Y}^2$.

Recall now that in conventional LR models with $\mu_{EW} \ll \mu_R \ll
\Lambda_{GUT}$ (or $\Lambda_{\rm Planck}$) one has to fine-tune two
gauge hierarchies: $\Lambda_{GUT}\dash \mu_R$ and $\mu_R\dash \mu_{EW}$.
We have a similar situation here: to achieve $\mu_{EW} \ll \mu_R
\ll \Lambda$ one has to fine-tune two Yukawa couplings, $Y_5^2$ and
$\tilde{Y}^2$. Tuning of $Y_5^2$ allows for the hierarchy
$\mu_R^2 \ll \Lambda^2$; one then needs to adjust
$\tilde{Y}^2$ (or $\mu_1^2$) to achieve $\mu_{EW}^2 \ll \mu_R^2$ through
eq.~(\ref{ratio}).

Since $\lambda_5$ only contains relatively small gauge couplings while
$Y_5 \sim \cal{O}$$(1)$, we typically have $|\lambda_5|/2\lambda_1
\sim 10^{-2}$. Thus, if there is no significant cancellation between the
two terms in (\ref{ratio}), one obtains a right--handed scale of the order of
a few $TeV$. Unfortunately, such a low LR scale scenario is not viable.
As we shall see below, the squared masses of two Higgs bosons in our
model become negative (i.e. the vacuum becomes unstable) unless
$v_R\,\gtap \,20~TeV$. This requires some cancellation\footnote{Notice
that this does not increase the number of the parameters to be tuned but
just shifts the value to which one of them should be adjusted.} in
eq.~(\ref{ratio}), and then the right-handed scale $v_R\sim \mu_R$ can in
principle lie anywhere between a few tens of $TeV$ and $\Lambda$.
However, if one prefers ``minimal cancellation'' in eq.~(\ref{ratio}),
by about two orders of magnitude or so, one would arrive at a value of
$v_R$ around $20~TeV$. In any case it is interesting that the partial
cancellation of the two terms in (\ref{ratio}) implies $\mu_1^2<0$,
i.e. that $\tilde{Y}^2$ must be below its critical value. This means that
$\tilde{M}_1^2$ never becomes negative. In fact it is the $\tilde{M}_0^2$
term, responsible for parity breakdown, that also drives the VEV of the
bi-doublet. It follows from the condition $\partial V_{\mbox{eff}}/\partial
\kappa=0$ that the effective driving term for $\kappa$ is $\tilde{M}_1^2+
\lambda_5 v_R^2$ in our model; it may become negative for large enough
$v_R^2$ even if $\tilde{M}_1^2$ is positive (remember that $\lambda_5<0$).
Thus we have a tumbling scenario where the breakdown of parity and $SU(2)_R$
occurring at the scale $\mu_R$ causes the breakdown of the \EW symmetry at a
lower scale $\mu_{EW}$.

To calculate physical observables one should first rescale the Higgs fields
so as to absorb the $Z$ factors in eq.~(\ref{Leff}) into the definitions of the
scalar fields and bring their kinetic terms into the canonical form.
This amounts to dividing the (mass)$^2$ terms by the corresponding $Z$
factors, Yukawa couplings by $\sqrt{Z}$ and multiplying the scalar fields
and their VEVs by $\sqrt{Z}$. Renormalization factors of the quartic
couplings depend on the scalars involved and can be readily read off from
the effective potential. We will use hats $(\,\hat{ }\,)$ to denote
quantities in the new normalisation.

As we already pointed out, the minimization of the effective Higgs potential
gives $\sigma_0=0=v_L$. This means that the entries $\beta$ and $\tilde{\mu}$
in the neutrino mass matrix (\ref{numass1}) are zero. As a result we have an
exactly massless neutrino eigenstate and two heavy Majorana neutrinos with
degenerate masses $\sqrt{M^{2}+m_{D}^{2}}$ and opposite $CP$--parities
which combine to form a heavy Dirac neutrino. Since $m_D \ll M$ the \EW
eigenstate $\nu_L\equiv \nu_{\tau}$ is predominantly the massless eigenstate
whereas the right--handed neutrino $\nu_R$ and the singlet fermion $S_L$
consists predominantly of the heavy eigenstates. As mentioned before, in the
simplified limit $\kappa'=0$ that we are mainly considering we have
$Y_2=0=Y_3$. This yields $m_t=Y_1\kappa$, $m_{\tau}=Y_4\kappa$ and
$m_b=m_D=0$. Vanishing Dirac neutrino mass $m_D$ implies absence of neutrino
mixing, and the heavy neutrino mass is now $M=Y_{5}v_{R}$. From
eqs.~(\ref{fmass}) and (\ref{zfactors}) and the definition of the
renormalized Yukawa couplings one can readily find
\beq
\hat{\kappa}^2=(174~GeV)^{2}
= N_c m_{t}^{2}\(1+\frac{Y_{4}^{2}}{N_{c}Y_{1}^{2}}\)\sixtpi\lnlambda
\approx  \frac{m_{t}^{2}N_c}{16\pi^2}\lnlambda
\equiv m_{t}^{2} N_c l_0~.
\label{mtopbubble}
\eeq
Here $\hat{\kappa}$ (or $\sqrt{\hat{\kappa}^2+\hat{\kappa}'^2}$ for
$\kappa'\neq 0$) should be identified with the usual \EW VEV.
This expression coincides with the one derived in the bubble approximation
by BHL \cite{BHL}. Eq.~(\ref{mtopbubble}) gives the top quark mass in terms
of the known \EW VEV and the scale of new physics $\Lambda$.
For example, for $\Lambda=10^{15}~GeV$ one finds $m_t\simeq 165~GeV$.
However, this result is limited to the bubble approximation, and the \RG
improved result for $\kappa'=0$ turns out to be
substantially higher\footnote{The \RG
improved values of $m_t$ will be viable for appropriate values of
$\tan\beta \equiv \kappa/\kappa'$ which in fact is a free parameter in
our model depending on the ratio $Y_4/Y_3$. In the limit
$\tan\beta \to \infty$ one obtains too high a top mass, e.g. $m_t=233~GeV$
for $\Lambda=10^{15}~GeV$,
whereas for $\tan\beta=(2.1-2.8)$, $\Lambda=10^{15}~GeV$ and
$\mu_R=10^7~GeV$ one finds $m_t=(168-192)~GeV$~\cite{ALSV2}.}
\cite{ALSV2}. Notice that $m_t \approx 180~GeV$, which is the central value
of the Fermilab results \cite{CDF,D0}, would mean $l_0 \approx 1/3$. Similar
considerations lead to the following relation between the right-handed VEV
$v_R$, the heavy neutrino mass $M$ and the scale $\Lambda$:
\beq
\hat{\vr2} =  M^{2}\sixtpi\lnlambda~ = M^2 \cdot l_0~.
\label{Vr2}
\eeq
Note that $\mu \approx m_t$ is understood in eq.~(\ref{mtopbubble}),
whereas $\mu \approx M$ in eq.~(\ref{Vr2}). However, we assume
$m_t,M\ll \Lambda$ and $M/m_t \ll \Lambda/M$  throughout this paper,
therefore $\ln\frac{\Lambda^2}{m_t^2} \approx \ln\frac{\Lambda^2}
{M^2}$, i.e. the logarithmic factor $l_0$ is universal. From
eqs.~(\ref{mtopbubble}) and (\ref{Vr2}) one thus finds
\beq
\frac{\hat{\vr2}}{M^2}\approx \frac{1}{3}\frac{\hat{\kappa}^2}{m_{t}^2}~.
\label{VrtoM}
\eeq
The mass of the $\tau$ lepton is not predicted in our model since it is
only weakly coupled to the bi-doublet; it is given by $m_{\tau} = (Y_4/Y_1)
m_t$ and can be adjusted to a desirable value by choosing the proper
magnitude of the ratio $Y_4/Y_1$, or $G_3/G_1$.

The composite Higgs bosons in our model include the would-be \GBs
$G_1^\pm \approx \chi_R^\pm$ (eaten by $W_R^\pm$), $G_2^\pm = \phi_1^\pm$
(eaten by $W_L^\pm$), $G_1^0 = \chi_{Ri}^0$ (eaten by $Z_R$) and  $G_2^0 =
\phi_{1i}^0$ (eaten by $Z_L$). The physical Higgs boson sector of the model
contains two $CP$--even neutral scalars $H_1^0 \approx \chi_{Rr}^0$ and
$H_2^0 \approx \phi_{1r}^0$ with the masses
\bea
M_{H_1^0}^2 & \simeq & 4M^{2}\[1-\frac{3}{16}
 \(3g^{4}+2g^{2}g'^{2}+g'^{4}\)l_0^2\]\approx 4M^2 \;,\label{MH10}\\
M_{H_2^0}^2 & \simeq & 4 m_t^2 \(1-\frac{m_{\tau}^{2}}{3 m_{t}^{2}} -
\frac{9}{4}g^4 l_0^2 \) \approx 4 m_{t}^{2}\;, \label{MH20}
\eea
which are directly related to the two steps of symmetry breaking,
$SU(2)_R\times U(1)_{B-L}\rightarrow U(1)_Y$ and
$SU(2)_L\times U(1)_Y    \rightarrow U(1)_{em}$. The mass of the
scalar $H_2^0$, which is the analog of the \SM Higgs boson
[eq.~(\ref{MH20})], essentially coincides with the one obtained in the
bubble approximation by BHL \cite{BHL}. This just reflects the fact that
this boson is the $t\bar{t}$ bound state with a mass of  $\approx 2 m_t$.
Analogously, the mass of the heavy $CP$--even scalar $H_1^0 \approx
\chi_{Rr}^0$ is approximately $2M$ since it is a bound state of heavy
neutrinos.

Further, there are the charged Higgs bosons $H_3^\pm \approx \phi_2^\pm$ with
their neutral $CP$--even and $CP$--odd partners $H_{3r}^0 =\phi_{2r}^0 $ and
$H_{3i}^0 =\phi_{2i}^0\;$, and finally the $\chi_L$-fields
$H_4^\pm = \chi_L^\pm\;$, $H_{4r}^0=\chi_{Lr}^0$  and
$H_{4i}^0=\chi_{Li}^0$ with the masses
\bea
M_{H_3^\pm}^2 &\approx&  \frac{2}{3}M^{2}\frac{m_{\tau}^{2}}{m_{t}^{2}}\;,
\label{MH3pm} \\
M_{H_{3r}^0}^2 &=& M_{H_{3i}^0}^2 \approx \frac{2}{3}M^{2}
\frac{m_{\tau}^{2}}{m_{t}^{2}} - \frac{1}{2} M_{H_2^0}^2~\;,\label{MH30}\\
M_{H_4^\pm}^2 &=& \frac{3}{8}\(3g_2^4+2g_2^2g_1^2\)l_0^2\,M^2+2 m_{\tau}^2
\label{MH4pm} \\
M_{H_{4r}^0}^2&=& M_{H_{4i}^0}^2 =
\frac{3}{8}\(3g_2^4+2g_2^2g_1^2\)l_0^2\,M^2~. \label{MH40}
\eea
In conventional LR models only one scalar, which is the analog of the \SM
Higgs boson, is light (at the \EW scale), all the others have their masses
of the order of the right--handed scale $M$ \cite{LR1,LR3,LR,LRdesh}. In
our case, the masses of those scalars are also proportional to $M$, but all
of them except the mass of $H_1^0$ have some suppression factors. The mass of
the charged scalars $H_3^\pm \approx \phi_2^\pm$ is suppressed by the factor
$m_{\tau}/m_t$ and is therefore of the order $10^{-2} M$. The masses of the
neutral $H_{3r}^0$ and $H_{3i}^0$ are even smaller; they are related to the
masses of of the charged $H_3^\pm$ and the \SM Higgs $H_2^0$ by
eq.~(\ref{MH30}). From the vacuum stability condition $M_{H_{3}^0}^2>0$ one
thus obtains an upper limit on the \SM Higgs boson mass $M_{H_2^0}$ (for a
given $M$) or a lower limit on the right--handed mass $M$ (for a given
$M_{H_2^0}$). For example, for $M_{H_2^0} \approx 60~GeV$ we find
$M\gtap 5~TeV$. However, since in the top condensate approach the \SM Higgs
mass is $\sim 2m_t$ (or $\sim m_t$ after the \RG improvement), a stronger
bound on the right handed gauge symmetry breaking scale of about $20~TeV$
results.

The masses of the $\chi_L$ scalars [eqs.~(\ref{MH4pm}),(\ref{MH40})]
vanish in the limit $(\lambda_2-\lambda_1) \to 0$ (i.e. $g_2 \to 0$) and
$m_\tau \to 0$. This fact has a simple interpretation. In the limit
$\lambda_2=\lambda_1$  (which corresponds to the fermion-bubble level)
the ($\chi_L,\chi_R$) sector of the effective Higgs potential
[eq.~(\ref{Veff})] depends on $\chi_L$ and $\chi_R$ only through the
combination $(\chi_L^\dagger \chi_L+\chi_R^\dagger \chi_R)$. This means
that the potential has a global $SU(4)$ symmetry which is larger than the
initial $SU(2)_L\times SU(2)_R\times U(1)_{B-L}$ symmetry. After $\chi_R^0$
gets a non-vanishing VEV $v_R$, the symmetry is broken down to $SU(3)$,
resulting in $15 - 8=7$ \GBsp Three of them ($\chi_R^\pm$ and $\Im \chi_R^0$)
are eaten by the $SU(2)_R$ gauge bosons $W_R^\pm$ and $Z_R$, and the
remaining four ($\chi_L^\pm$, $\Re \chi_L^0$ and $\Im \chi_R^0$) are
physical massless \GBsp The $SU(4)$ symmetry is broken by the $\phi$--dependent
terms in the effective potential and by $SU(2)$ gauge interactions.
As a result, $\chi_L^\pm$, $\Re \chi_L^0$ and $\Im \chi_L^0$ acquire small
masses and become pseudo--\GBsp In fact, the origin of this approximate
$SU(4)$ symmetry can be traced back to the 4-f operators of eq.~(\ref{L4f}).
It is an accidental symmetry resulting from the gauge invariance and parity
symmetry of the $G_7$ term. Note that no such symmetry occurs in conventional
LR models.

Finally, we would like to comment on the approximation $\kappa'=0$ which
we have used. If we relax this condition, we will obtain non-vanishing
masses $m_b$ and $m_D$ (notice that the Yukawa couplings $Y_2$ and $Y_3$
will also be non-zero in this case). However, these masses are not predicted
in our model and can simply be adjusted to desirable values. The Dirac
neutrino mass $m_D$ is unknown and so remains a free parameter; however,
it must be smaller than $m_{\tau}$ in our model in order to satisfy the
vacuum stability condition $Y_4^2-Y_3^2>0$ \cite{ALSV2} which is equivalent
to $m_{\tau}^2-m_D^2>0$. For $\kappa' \ll \kappa$ our predictions for the
Higgs boson masses are only slightly modified. As our renormalization--group
analysis performed in \cite{ALSV2} shows, some interesting results emerge
for sizeable values of $\kappa'$. The Higgs boson masses and mass eigenstates
for the general case $\kappa' \ne 0$ can be found in \cite{ALSV2}.

In our model we have 9 input parameters (eight 4-f couplings $G_1,...,G_8$
and the scale of new physics $\Lambda$) in terms of which we calculate 16
physical observables (5 fermion masses, 8 Higgs boson masses and 3 VEVs
$\kappa$, $\kappa'$ and $v_R$), so there are $16-9=7$ predictions. In the
simplified case $\kappa'=0$ that we were mainly considering we have only
5 input parameters since $\kappa'=0$ requires $G_2=G_4=G_5=0$, $G_6 =
\sqrt{G_1G_3}$. At the same time we have only 13 non-trivial physical
observables since the bottom quark mass and Dirac neutrino mass vanish
identically in this case. This yields $13-5=8$ predictions.

To summarize, this is to our knowledge the first successful attempt to
break LR symmetry dynamically. We find a tumbling scenario where the breaking
of parity and $SU(2)_R$ eventually drives the breaking of the \EW symmetry.
The model gives a viable top quark mass value and exhibits a number of low
and intermediate scale Higgs bosons. Furthermore it predicts relations
between masses of various scalars and between fermion and Higgs boson masses
which are in principle testable. If the right--handed scale $\mu_R$ is of
the order of a few tens of $TeV$, the neutral $CP$--even and $CP$--odd
scalars $\phi_{2r}^0$ and $\phi_{2i}^0$ can be even lighter than the \EW
Higgs boson. In fact, they can be as light as $\sim 50~GeV$ and so might
be observable at LEP2. Such light $\phi_{2r}^0$ and $\phi_{2i}^0$ can also
provide a positive contribution to
$R_b=\Gamma(Z\to b\bar{b})/\Gamma(Z\to hadrons)$ \cite{Denner1} which is
necessary to account for the discrepancy between the LEP observations and the
\SM predictions.

We are grateful to Zurab~Berezhiani, Darwin~Chang, Richard~Dawid,
Rabi~Mohapatra, Ulrich~Nierste, Graham~Ross, Goran~Senjanovi\'{c} and
Alexei~Smirnov for useful discussions. Thanks are also due to
Stefano~Ranfone, who participated in the earlier stage of the work.
This work was supported in part by the Spanish DGICYT under grants PB92-0084
and SAB93-0090 (E.A. and J.V.) and by the German DFG under contract
number Li519/2--1 (M.L. and E.S.).


\end{document}